\documentclass{elsart1p}
\usepackage{graphicx}
\usepackage{amssymb}
\begin{document}
\begin{frontmatter}
% use the corauthref command within \author for corresponding author
% footnotes;
\title{Prospects for New Physics in CP Violation and Rare Decays at LHCb}
\author{Pascal Perret\thanksref{behalf}}
\thanks[behalf]{on behalf of the LHCb Collaboration}
\address{Laboratoire de Physique Corpusculaire, Universit\'e Blaise Pascal - CNRS/IN2P3,
24 avenue des Landais, F 63 177 Aubi\`ere cedex, France.}
\ead{perret@in2p3.fr}
\begin{abstract}
% Text of abstract
LHCb is the heavy flavour precision experiment of the proton-proton Large Hadron Collider (LHC) at CERN. It will search for new physics in CP violation and rare decays and is ready for the start-up of the LHC. An overview of its physics program will be given, illustrated by few key examples:
measurements of the CKM angle $\gamma$, of the CP violating phase in $ B_s \overline{B_s}$ oscillations and searches for new physics in $B^0_d \to K^*\mu^+\mu^-$ and in $B^0_s \to \mu^+\mu^-$.
\end{abstract}
\end{frontmatter}
%
% main text
\section{Introduction}
\label{sec:Intro}
LHCb~\cite{jinst} is the dedicated $b$-physics experiment at the proton-proton LHC collider installed at CERN. It will benefit from the large $b \bar b$ cross section of $\sim$ 500 $\mu$b at a centre-of-mass energy of 14 TeV and from the correlated $b \bar b$ pair production that is sharply peaked forward-backward. This is why the LHCb detector is a single-arm forward spectrometer with an acceptance in $\theta \sim [15-300]$ mrad, 
corresponding to the rapidity range: $[1.9 - 4.9]$.
In addition all flavours of $b$-hadrons will be produced ($B_{d}$, $B_{u}$, $B_{s}$, $B_{c}$, $\Lambda_{b}$, ...).

LHCb will run at a limited luminosity of 2 $\times$ $10^{32}$
cm$^{-2}$s$^{-1}$, instead of the nominal LHC luminosity of $10^{34}$
cm$^{-2}$s$^{-1}$, obtained by not focusing the beam at the LHCb interaction point as much as ATLAS and CMS. This has been chosen to maximize the number of single interactions per
crossing in order to produce clean events and to facilitate the triggering
and reconstruction. This permits also to make radiation damage in the forward region
more manageable and to reach the design luminosity soon after the start-up.

The detector is fully installed and commissioned (except the first layer of the muon system, which is not needed at the start-up), including the first trigger level system. All sub-detectors have undergone the first time and space alignment with cosmic events and LHC beam induced particles produced this summer.

The unprecedented quantity of $b$-hadrons that will be produced will allow LHCb to provide stringent tests of the Standard Model (SM) and to improve significantly with respect to the present status our knowledge of the unitarity triangle and of the CKM matrix~\cite{CKM}.  
In addition, $b$-hadron decays occurring via penguin or box diagrams are numerous and are an important laboratory where new particles may show up.  
Thus $b$-physics measurements probe New Physics (NP). They are complementary to direct searches and should allow to understand the nature and flavour structure of possible NP.

A few key measurements which will be conducted by LHCb are addressed here (more details in~\cite{reop_tdr}).  
In particular the precision on the angle $\gamma$ is expected to improve dramatically. Also of the highest interest will be a precise measurement of the CP violating phase in $ B_s \overline{B_s}$ oscillations. In addition, a unique opportunity is provided for studying very rare decays where the effect of physics beyond the Standard Model might dominate. For instance, the decay $B_s \to \mu\mu$  is very sensitive to an extended Higgs sector and might be the first place that the effects of Supersymmetry are seen in data from the LHC. 

\section{$\gamma$ measurements}
The decay $B^0_s \to D^\pm_s K^\mp$ and its charge conjugate can
proceed through two tree  decay diagrams, the interference of which
gives access to the phase $\gamma +  \phi_s$, and hence to the CKM
angle $\gamma$ if $\phi_s$ is determined otherwise ($e.g.$ with 
$B_s \to J/\psi \phi$, see section~\ref{sect:phys}). Four time-dependent decay rates are
measured. There is a background from $B^0_s
\to D^-_s \pi^+$ decays, which has a branching fraction $\sim$20 times
higher than the signal channel, but is successfully controlled using the particle identification and good mass resolution of LHCb. 
About 6.2k signal events will be recorded in one year (2 fb$^{-1}$), giving a sensitivity
of 10$^{\circ}$ in $\gamma$. There are different ways to measure $\gamma$ at LHCb. Using other $B \to D K$ decays, a sensitivity of $\sim$5$^{\circ}$ can be reached with 2 fb$^{-1}$.

\section{$\phi_s$ measurements from $B_s \to J/\psi \phi$}
\label{sect:phys}
The $B^0_s$ mixing phase $\phi_s$ is very small in the SM ($\sim$ -0.04), but it can receive sizeable NP contributions through box diagrams. First measurements have been performed at the Tevatron~\cite{phisTevatron}, around 2.2 $\sigma$ away from the SM expectation but with very poor precision yet. The golden mode for measuring it at LHCb is via the time-dependent decay rate asymmetries of $B_s \to J/\psi \phi$. The branching ratio is (3.1 $\pm$ 1.1)$\times$$10^{-5}$.This analysis is complicated by the fact that the final state involves two vector mesons and so two orbital momentum states can occur. Therefore an angular analysis of the decay final states is needed to separate CP-even and CP-odd contributions. Experimentally,
good proper-time resolution is essential, but also good tagging of initial $B_s$ flavour, good mass resolution and good particle identification. A proper time resolution better than $<$39 fs, with a combined tagging power of 6.2\%, a mass resolution of 13 MeV and a kaon identification efficiency of 80\% for a mistag rate of 3\% should be achieved.

LHCb will reconstruct 114k events per year (with 2 fb$^{-1}$ of data taking), with S/B=2. This will allow a statistical precision of 0.03 on $\phi_s$ with low systematic uncertainties. Already at the start-up with 0.5 fb$^{-1}$, LHCb will reconstruct 28.5k events and should rapidly pin down whether there really is any sign of new physics.

\section{New physics in $B^0_d \to K^*\mu^+\mu^-$}
\label{sec:NewPhysics}
Flavour-changing neutral current decays, such as $B^0_d \to K^*\mu^+\mu^-$, are also very interesting to probe NP. While in the SM it occurs via a suppressed $b \to s$ electroweak penguin diagram, NP diagrams could contribute at the same level. Among the observables which can be built, the Forward-Backward asymmetry A$_{\rm FB}$ in the $\mu \mu$ rest frame is of interest, and in particular the invariant $\mu \mu$ mass value where A$_{\rm FB}$ reach its zero value ($s_0$). The SM model prediction is precise, while NP (and SUSY) could give a sizeable deviation. Some first measurements have been conducted at B factories, but are statistically limited.

LHCb is developing a selection with an expected yield of 7.2k events/2fb$^{-1}$ with a B/S of 0.2; to be compared with the 450 events expected from the B factories with a 2 ab$^{-1}$ integrated luminosity. A simple linear fit suggests a precision on $s_0$ of 0.46 GeV$^2$ with 2 fb$^{-1}$.

\section{New physics in $B^0_s \to \mu^+\mu^-$}
The $B^0_s \to \mu^+\mu^-$ decay is predicted to be very rare in the SM (${\cal B} = (3.4 \pm 0.5) \times 10^{-9}$) since it involves flavour-changing neutral currents and experiences a large helicity suppression, but is sensitive to NP and could be strongly enhanced in SUSY. The best current limit is achieved by CDF (${\cal B} < 4.7 \times 10^{-8} @ 90\%$ CL)~\cite{CDF_bsmu}.

After a first selection mainly based on lifetime information, a classification of each event will be performed in a 3D phase space built with a geometrical likelihood (using topological information), mass and particle identification information. For each bin the expected number of events for the signal and for the signal-plus-background hypothesis will be computed and used to measure the branching ratio (BR), or place a limit. 

The LHCb experiment has the potential of measuring a BR of 9(15)$\times 10^{-9}$ at 3(5)$\sigma$ level with 0.1 fb$^{-1}$ of data (corresponding to the Tevatron sensitivity with their complete dataset) and 5(9)$\times 10^{-9}$ with 0.5 fb$^{-1}$. The SM BR can be measured with a 3(5)$\sigma$  evidence (observation) with 2(6) fb$^{-1}$.

% \appendix
%
% \section{}
% \label{}
%

%

\begin{thebibliography}{00}
% notes:
% \bibitem{label2}
% If there is a note, it should come last:
% \bibitem{label3} \note
% \end{subbibitems}
%

\bibitem{jinst} The LHCb Collaboration, %{\it "The LHCb detector at the LHC"}, 
Journal of Instrumentation, 3 (2008) S08005.
%
\bibitem{CKM} M.~Kobayashi and T~.~Maskawa, Prog.
Theor.\ Phys.\ {\bf 49} (1973) 652.
%
\bibitem{reop_tdr}  The LHCb Collaboration, 
CERN LHCC 2003-30.
\bibitem{phisTevatron}
The CDF Collaboration,      Phys. Rev. Lett. {\bf 100}, 161802 (2008); \\
The D0 Collaboration,      Phys. Rev. Lett. {\bf 101}, 241801 (2008).
%
\bibitem{CDF_bsmu} The CDF Collaboration,    Phys. Rev. Lett. {\bf 100}, 101802 (2008).
%

\end{thebibliography}
\end{document}